\documentclass[10pt]{article}
\usepackage{graphicx}
\usepackage{amsmath}
\usepackage{amssymb}
\usepackage{caption2}
\begin{document}
\bibliographystyle{prsty}
\begin{center}
{\large {\bf \sc{  The  $B_c$-decays  $B_c^+ \to J/\psi\, \pi^+\pi^-\pi^+$, $\eta_c\, \pi^+\pi^-\pi^+ $  }}} \\[2mm]
Zhi-Gang Wang \footnote{E-mail,wangzgyiti@yahoo.com.cn.  }     \\
 Department of Physics, North China Electric Power University,
Baoding 071003, P. R. China
\end{center}

\begin{abstract}
In this article, we study the three-pion $B_c$-decays $B_c^+ \to J/\psi\, \pi^+\pi^-\pi^+$, $\eta_c\, \pi^+\pi^-\pi^+ $  with  dominance of the intermediate  axial-vector   meson $a_1(1260)$ and vector meson $\rho(770)$   in the $\pi^+\pi^-\pi^+$ and $\pi^+\pi^-$ invariant mass distributions respectively,
 and make predictions for the branching fractions and differential decay widths. The   ratio between the decays $B_c^+ \to J/\psi\, \pi^+\pi^-\pi^+$ and $B_c^+ \to J/\psi\, \pi^+ $ is compatible with the experimental data within uncertainties, other predictions can be confronted with the experimental data in the future at the LHCb.
\end{abstract}

 PACS number: 13.25.Hw

Key words: $B_c$-meson decays

\section{Introduction}

In 1998, the CDF collaboration  observed the pseudoscalar  $B^{\pm}_c$ mesons  through  the semileptonic decays $B_c^{\pm} \to J/\psi \ell^{\pm}X $ and $B_c^{\pm} \to J/\psi \ell^{\pm}\bar{\nu}_{\ell} $ in $p\bar{p}$ collisions at the energy $\sqrt{s}=1.8\,\rm{TeV}$ at the Fermilab Tevatron \cite{CDF1998}.
The bottom-charm  quarkonium states $B_c$  are of special interesting, the ground states and the excited states lying below the $BD$, $BD^*$, $B^*D$, $B^*D^*$ thresholds cannot annihilate into gluons, and therefore  are more stable than the corresponding heavy quarkonium states consist of the same flavor, and would have widths less than a hundred $\rm{KeV}$ \cite{GI}. The semileptonic decays $B_c^{\pm} \to J/\psi \ell^{\pm}\bar{\nu}_{\ell}$, $B_c^{+} \to J/\psi e^{+}\bar{\nu}_{e} $ were   used to measure the $B_c$ lifetime \cite{CDF-life,D0-life},
which is about a third as long as that of the   $B$ and $B_s$ mesons as both  the $b$ and $c$ quarks decay weakly.   In 2007, the  CDF collaboration  observed the $B_c^{\pm}$ mesons with an significance exceeds  $8\, \sigma$ through the decays $B_c^{\pm} \to J/\psi \pi^{\pm}$  in $p\bar{p}$ collisions at the energy  $\sqrt{s}=1.96\,\rm{TeV}$, and obtained the value  $m_{B_c}=(6275.6 \pm 2.9\pm 2.5) \, \rm{MeV}$ \cite{CDF2008}.  In 2008, the D0 collaboration reconstructed the decay modes $B_c^{\pm} \to J/\psi \pi^{\pm}$  and observed the  $B_c^{\pm}$ mesons  with an significance larger  than $5\,\sigma$, and obtained the value $m_{B_c}=(6300 \pm 14 \pm 5)\,\rm{ MeV}$ \cite{D02008}. Now the average value is $m_{B_c}=(6.277 \pm 0.006)\,\rm{GeV}$  from the Particle Data Group \cite{PDG}.
Recently, the LHCb collaboration  observed the decay $B_c^+ \to J/\psi \pi^+\pi^-\pi^+$   for the first time  using $0.8 \,\rm{fb}^{-1}$ of the $pp$ collisions at $\sqrt{s}=7 \,\rm{TeV}$, the measured ratio of branching fractions is
\begin{eqnarray}
\frac{Br(B_c^+ \to J/\psi \pi^+\pi^-\pi^+)}{Br(B_c^+ \to J/\psi \pi^+)} &=& 2.41\pm0.30\pm0.33 \, ,
\end{eqnarray}
 where the first uncertainty is statistical and the second is systematic \cite{LHCb-Bc}.

 The hadronic decays $B_c^+ \to J/\psi \,\pi^+\pi^-\pi^+,\,J/\psi \,\pi^+ $ take place through the  weak $b$-quark decays
$b\to cW^{*}\to c\bar{u}d$, which are analogous  with the $\tau$-lepton
decays $\tau\to \nu_{\tau}W^{*}\to \nu_{\tau}\bar{u}d$. We can use the existing experimental data on the $\tau$-lepton decays
to obtain reliable prediction for the  $B^+_{c}\to J/\psi\, \pi^+\pi^-\pi^+$ branching
fraction with the method of spectral functions,
\begin{eqnarray}
\int d\Phi(W^{*}\to\pi^+\pi^-\pi^+)\epsilon_{\mu}\epsilon_{\nu}^{*} & = &(q_{\mu}q_{\nu}-q^2g_{\mu\nu})\rho_{T}(q^2)+q_{\mu}q_{\nu}\rho_{L}(q^2)\, ,
\end{eqnarray}
where the $d\Phi(W^{*}\to\pi^+\pi^-\pi^+)$ is the Lorentz-invariant three-body phase factor, the $\epsilon_\mu$ is the effective polarization vector of the
virtual $W$-boson,
the spectral functions $\rho_{T}(q^2)$ and $\rho_{L}(q^2)$
are universal and can be determined by theoretical and experimental
analysis of the   $\tau\to\nu_{\tau} \pi^+\pi^-\pi^+$ decays \cite{Likhoded-3pi,Berezhnoy-3pi,Berezhnoy-3pi-2,Rakitin-3pi}.
The spectral function $\rho_{L}(q^2)$ is negligible according to   conservation of vector current and partial conservation of axial-vector current. The explicit
expression of the spectral function $\rho_{T}(q^2)$ can be fitted to the experimental data or   calculated by some phenomenological models,
 the spectral function $\rho_{T}(q^2)$ is always  saturated by exchange of the intermediate axial-vector meson $a_1(1260)$ with an special ansatz for  the
 vertexes  $a_1\rho\pi$ and $\rho\pi\pi$ based on some phenomenological quark models  \cite{Kuhn-3pi-ZPC,Li-3pi-PRD,Ivanov-3pi-ZPC,Kuhn-3pi-PLB}.
In this article, we intend to study  the decays $B_c^+ \to J/\psi \pi^+\pi^-\pi^+$, $ \eta_c \pi^+\pi^-\pi^+ $, $J/\psi \pi^+ $, $ \eta_c  \pi^+ $   with the phenomenological Lagrangians, and calculate the Feynman diagrams directly. The decays $B_c^+ \to   \eta_c\, \pi^+\pi^-\pi^+$, $\eta_c \pi^+ $ have not been observed yet, but they are expected to be observed in the future at the  Large Hadron Collider (LHC). The $B_c$ decays will be studied and their branching fractions will be determined
in the LHCb experiments.

In Ref.\cite{a1-rho-pi}, Lichard and Juran perform detailed analysis of the vertex $a_1\rho \pi$ with the following Lagrangian,
\begin{eqnarray}
\cal{L}&=&\frac{g_{a\rho\pi}}{\sqrt{2}}\left\{ cos\theta \left[(\partial^\mu \rho^{0\nu}-\partial^\nu \rho^{0\mu})a_\mu^{-}\partial_\nu \pi^{+}\right]
-sin\theta\left[(\partial^\mu \rho^{0\nu}-\partial^\nu \rho^{0\mu})\partial_\mu a_\nu^{-} \pi^{+}\right]+\cdots \right\} \, ,
\end{eqnarray}
considering  the decays $a_1(1260)\to \rho \pi$, where the momenta of the $\rho$ and $\pi$ mesons in center-of-mass frame
of the initial $a_1(1260)$ meson are about $0.37\,\rm{GeV}$. Such  a Lagrangian maybe lead to amplified  amplitude for the decay $B_c^+ \to J/\psi\, \pi^+\pi^-\pi^+$ as the sub-amplitude $a^+_1(1260) \to \rho\pi^+$ can be accounted as  $g_{a\rho\pi}l_\rho \cdot k_\pi$ or $g_{a\rho\pi}l_\rho \cdot q_a$, where the momenta  $q_a$, $l_\rho$ and $k_{\pi}$ are large, we have to introduce form-factors to parameterize the
off-shell effects.
 In the decays $\tau \to a_1(1260)\,\nu_{\tau}$ and $B_c^+\to J/\psi\, a^+_1(1260)$, the momenta of the $a_1(1260)$ meson
in center-of-mass frame of the initial  particles are about $0.45\,\rm{GeV}$ and $2.16\,\rm{GeV}$, respectively.
 On the other hand, we know that the decays $a_1(1260)\to \rho \pi$ are $S$-wave dominated \cite{PDG},  we prefer
the simple  Lagrangian,
\begin{eqnarray}
{\cal{L}}_{a\rho\pi}&=&g_{a\rho\pi} a_1^\mu \rho_\mu \pi \, ,
\end{eqnarray}
in this article. Furthermore, we use the Lagrangian,
\begin{eqnarray}
\cal{L}_{\rho\pi\pi}&=&-ig_{\rho\pi\pi} \rho^{0\mu}\pi^-( \overrightarrow{\partial}_\mu-\overleftarrow{\partial}_\mu) \pi^+ \, ,
\end{eqnarray}
to study the vertex $\rho\pi\pi$ \cite{rho-pi-pi}.

The article is arranged as follows:  we derive the decay widths of the processes $B_c^+ \to J/\psi\,\pi^+\pi^-\pi^+$, $\eta_c\,\pi^+\pi^-\pi^+$, $ J/\psi\,\pi^+$, $\eta_c\,\pi^+$   in Sect.2; in Sect.3, we present the numerical results and discussions; and Sect.4 is reserved for our
conclusions.

\section{ Decay widths of the processes  $B_c^+ \to J/\psi\pi^+\pi^-\pi^+,\,\eta_c\pi^+\pi^-\pi^+$ }
The hadronic decays $B_c^+\to J/\psi\pi^+$,  $\eta_c\pi^+$,  $J/\psi\,\pi^+\pi^-\pi^+$ and $ \eta_c\, \pi^+\pi^-\pi^+$
can be described by the effective Hamiltonian,
 \begin{eqnarray}
 {\cal H}_{\rm eff}&=&\frac{G_F}{\sqrt{2}}V_{cb}V^*_{ud} C_1(\mu)\bar{c}\gamma_{\alpha}(1-\gamma_5)b \,\bar{d}\gamma^{\alpha}(1-\gamma_5)u+h.c. \, ,
 \end{eqnarray}
where the $V_{cb}$, $V_{ud}$ are the CKM matrix elements, the $G_F$ is the
Fermi constant, and the $C_1(\mu)$ is the Wilson coefficient defined at an special energy scale, $C_1(m_b)=1.14$ \cite{ff-QCDSR}.  In the following, we write down the definitions for the weak
form-factors $F_1(q^2)$, $F_0(q^2)$,  $A_1(q^2)$, $A_2(q^2)$, $A_3(q^2)$, $A_0(q^2)$ and $V(q^2)$  for the current $J_\mu=\bar{c}\gamma_{\mu}(1-\gamma_5)b$ sandwiched between the $B_c$ and the $\eta_c$, $J/\psi$ \cite{WSB},
\begin{eqnarray}
\langle {\eta_c}(p)|J_\mu(0) |B_c(P)\rangle&=& \left[(P+p)_\mu-\left(m_{B_c}^2-m_{J/\psi}^2\right)\frac{q_\mu}{q^2} \right]F_1(q^2)+\left(m_{B_c}^2-m_{J/\psi}^2\right)\frac{q_\mu}{q^2}F_0(q^2) \, , \nonumber\\
&=& (P+p)_\mu F_+(q^2)+ q_\mu F_-(q^2) \, ,  \\
\langle {J/\psi}(p)|J_\mu(0) |B_c(P)\rangle&=&i\left\{
-\epsilon_\mu^*(m_{B_c}+m_{J/\psi})A_1(q^2)+\epsilon^* \cdot P (P+p)_\mu
 \frac{A_2(q^2)}{m_{B_c}+m_{J/\psi}} + \right. \nonumber\\
&&\left.2m_{J/\psi}\epsilon^* \cdot P\frac{q_\mu}{q^2}\left[A_3(q^2)-A_0(q^2)\right]
-i\epsilon_{\mu\nu\alpha\beta}\epsilon^{*\nu} q^\alpha (P+p)^\beta  \frac{V(q^2)}{m_{B_c}+m_{J/\psi}}
\right\} \, , \nonumber\\
&=&i\left\{
-\epsilon_\mu^*(m_{B_c}+m_{J/\psi})A_1(q^2)+\epsilon^* \cdot P (P+p)_\mu
 \frac{A_+(q^2)}{m_{B_c}+m_{J/\psi}} + \right. \nonumber\\
&&\left.\epsilon^* \cdot P q_\mu
 \frac{A_-(q^2)}{m_{B_c}+m_{J/\psi}}
-i\epsilon_{\mu\nu\alpha\beta}\epsilon^{*\nu} q^\alpha (P+p)^\beta  \frac{V(q^2)}{m_{B_c}+m_{J/\psi}}
\right\} \, ,
\end{eqnarray}
where
\begin{eqnarray}
F_+(q^2)&=&F_1(q^2)\, , \nonumber \\
F_-(q^2)&=&\frac{\left[F_0(q^2)-F_1(q^2)\right]\left(m_{B_c}^2-m_{J/\psi}^2\right)}{q^2}\, , \nonumber \\
A_3(q^2)&=&\frac{m_{B_c}+m_{J/\psi}}{2m_{J/\psi}}A_1(q^2)-\frac{m_{B_c}-m_{J/\psi}}{2m_{J/\psi}}A_2(q^2)\, , \nonumber \\
A_+(q^2)&=&A_2(q^2)\, , \nonumber \\
A_-(q^2)&=&\frac{2m_{J/\psi}(m_{B_c}+m_{J/\psi})}{q^2}\left[ A_3(q^2)-A_0(q^2)\right]\, ,
\end{eqnarray}
and $V_0(0)=V_3(0)$, and the $\epsilon_\mu$ is the polarization vector
of the vector meson $J/\psi$.

There have been several approaches to calculate those weak form-factors, such as the QCD sum rules \cite{ff-QCDSR,KLO-QCDSR,Colangelo-QCDSR}, the
light-cone QCD sum rules \cite{HuangZuo}, the relativistic quark models \cite{ff-RQM1,ff-RQM2,IKS-RQM,Nobes-RQM}, the light-front quark models \cite{ff-LFQM,Choi-LFQM}, the non-relativistic quark models \cite{Hernandez-NRQM,Dhir-NRQM},
perturbative QCD \cite{Sun-pQCD}, etc. In this article, we take the typical values from    the QCD sum rules  \cite{ff-QCDSR}, the relativistic quark models \cite{ff-RQM1,ff-RQM2} and the light-front quark models  \cite{ff-LFQM}, and refer them as QCDSR, RQM1, RQM2 and LFQM, respectively.
In calculations, we  use the following definition for the decay constant $f_a$ of the axial-vector meson $a_1(1260)$,
\begin{eqnarray}
\langle 0|\bar{u}(0)\gamma_\mu \gamma_5 d(0)|a_1(1260)\rangle &=&f_am_a\varepsilon_\mu \, ,
\end{eqnarray}
where the $\varepsilon_\mu$ is the polarization vector.

We take into account the effective Hamiltonian  ${\cal H}_{\rm eff}$,  the weak form-factors and the Lagrangians ${\cal{L}}_{a\rho\pi}$, $\cal{L}_{\rho\pi\pi}$
to obtain the amplitudes $T_{J/\psi\pi^+\pi^-\pi^+}$, $T_{\eta_c\pi^+\pi^-\pi^+}$ of the processes $B_c^+\to J/\psi \, \pi^+\pi^-\pi^+$, $\eta_c \, \pi^+\pi^-\pi^+$,
\begin{eqnarray}
T_j&=&\frac{G_F}{\sqrt{2}}V_{cb}V^*_{ud}C_1(m_b) f_am_a\widetilde{T}_j \, , \nonumber\\
\widetilde{T}_j&=&\widetilde{T}^1_j+\widetilde{T}^2_j\, ,
\end{eqnarray}
where $j=J/\psi\,\pi^+\pi^-\pi^+$, $\eta_c\,\pi^+\pi^-\pi^+$,
\begin{eqnarray}
\widetilde{T}^1_{J/\psi \pi^+\pi^-\pi^+}&=&i\left\{-\epsilon_\mu^*(m_{B_c}+m_{J/\psi})A_1(q^2)+\epsilon^* \cdot P (P+p)_\mu \frac{A_+(q^2)}{m_{B_c}+m_{J/\psi}} +\epsilon^* \cdot P q_\mu  \frac{A_-(q^2)}{m_{B_c}+m_{J/\psi}} \right. \nonumber\\
&&\left.-i\epsilon_{\mu\lambda\tau\sigma}\epsilon^{*\lambda} q^\tau (P+p)^\sigma  \frac{V(q^2)}{m_{B_c}+m_{J/\psi}}\right\} \frac{i}{q^2-m_a^2+i\sqrt{q^2}\Gamma_a(q^2)}\left( -g^{\mu\nu}+\frac{q^\mu q^\nu}{q^2}\right)ig_{a\rho\pi} \nonumber\\
&&\frac{i}{l^2-m_\rho^2+i\sqrt{l^2}\Gamma_\rho(l^2)}\left( -g_{\nu\alpha}+\frac{l_\nu l_\alpha}{l^2}\right)ig_{\rho\pi\pi}(t-r)^\alpha \, ,\\
 \widetilde{T}^2_{J/\psi \pi^+\pi^-\pi^+}&=&\widetilde{T}^1_{J/\psi \pi^+\pi^-\pi^+} (\pi^+(r)\leftrightarrow\pi^+(k)) \, , \\
 \widetilde{T}^1_{\eta_c \pi^+\pi^-\pi^+}&=&\left\{ (P+p)_\mu F_{+}(q^2)+q_\mu F_{-}(q^2) \right\} \frac{i}{q^2-m_a^2+i\sqrt{q^2}\Gamma_a(q^2)}\left( -g^{\mu\nu}+\frac{q^\mu q^\nu}{q^2}\right)ig_{a\rho\pi} \nonumber\\
&&\frac{i}{l^2-m_\rho^2+i\sqrt{l^2}\Gamma_\rho(l^2)}\left( -g_{\nu\alpha}+\frac{l_\nu l_\alpha}{l^2}\right)ig_{\rho\pi\pi}(t-r)^\alpha \, ,\\
 \widetilde{T}^2_{\eta_c \pi^+\pi^-\pi^+}&=&\widetilde{T}^1_{\eta_c \pi^+\pi^-\pi^+} (\pi^+(r)\leftrightarrow\pi^+(k)) \, , \\
   \Gamma_a(q^2)&=&\Gamma_a \frac{m_a^2}{q^2}\left( \frac{q^2-9m_\pi^2}{m^2_a-9m_\pi^2}\right)^{\frac{3}{2}} \, ,\nonumber \\
  \Gamma_\rho(l^2)&=&\Gamma_\rho \frac{m_\rho^2}{l^2}\left( \frac{l^2-4m_\pi^2}{m^2_\rho-4m_\pi^2}\right)^{\frac{3}{2}} \, ,
  \end{eqnarray}
then obtain the differential  decay widths
\begin{eqnarray}
d\Gamma_j&=&\frac{|T_j|^2}{4m_{B_c}}\frac{dq^2}{2\pi}\frac{dl^2}{2\pi}d\Phi(P\to q,p)d\Phi(q\to l,k)d\Phi(l\to r,t) \, ,
\end{eqnarray}
where the $d\Phi(P\to q,p)$, $d\Phi(q\to l,k)$, $d\Phi(l\to r,t)$ are the two-body phase factors defined  analogously, for example,
\begin{eqnarray}
d\Phi(P\to q,p)&=&(2\pi)^4\delta^4(P-q-p)\frac{d^3\vec{p}}{(2\pi)^3 2p_0}\frac{d^3\vec{q}}{(2\pi)^3 2q_0} \, .
\end{eqnarray}
The decay widths of the processes $B_c^+\to J/\psi \pi^+$, $\eta_c\pi^+$ can be calculated   straightforward using the
effective Hamiltonian  ${\cal H}_{\rm eff}$ and the weak form-factors, the explicit expressions are neglected for simplicity.
\section{Numerical results and discussions}
The input parameters are taken as $G_F=1.166364\times 10^{-5}\,\rm{GeV}^{-2}$, $V_{ud}=0.97425$, $V_{cb}=40.6\times 10^{-3}$, $m_{\pi}=139.57\,\rm{MeV}$,
 $m_{\rho}=775.49\,\rm{MeV}$, $\Gamma_\rho=146.2\,\rm{MeV}$, $m_{B_c}=6.277 \,\rm{GeV}$, $\tau_{B_c}=0.45\times 10^{-12}\,s$ from the Particle Data Group \cite{PDG}, $m_{a}=1255\,\rm{MeV}$, $\Gamma_a=367\,\rm{MeV}$ from the COMPASS collaboration \cite{Compass}, $g_{\rho\pi\pi}=6.05$ from the decay $\rho \to \pi\pi$ \cite{rho-pi-pi},    $g_{a\rho\pi}=3.37$  from the decay $a_{1}(1260)\to\rho\pi$ \cite{PDG},
 $f_a=0.24\,\rm{GeV}$   from the QCD sum rules \cite{Yang-fa}.

 We obtain the branching fractions of the decays $B^+_{c}\to J/\psi\, \pi^+$,
 $ \eta_c\, \pi^+$, $ J/\psi \, \pi^+\pi^-\pi^+$ and   $ \eta_c\, \pi^+\pi^-\pi^+$ with the
  typical values of the weak form-factors from  the QCD sum rules  \cite{ff-QCDSR}, the relativistic quark models \cite{ff-RQM1,ff-RQM2},
  and the light-front quark models  \cite{ff-LFQM}. The form-factors  in Refs.\cite{ff-QCDSR,ff-RQM1,ff-RQM2} are fitted to an single pole form,
  \begin{eqnarray}
  f(q^2)&=&\frac{f(0)}{1-q^2/m^2_{fit}} \, ,
  \end{eqnarray}
  while     the form-factors in Ref.\cite{ff-LFQM}   are fitted to an  exponential form,
  \begin{eqnarray}
  f(q^2)&=&f(0)\exp\left(c_1 q^2+c_2 q^4\right)\, ,
  \end{eqnarray}
   where the $f(q^2)$ denote the weak form-factors, the $m_{fit}$, $c_1$, $c_2$ are fitted parameters. The numerical values are presented explicitly in Table 1.
   \begin{table}
\begin{center}
\begin{tabular}{|c|c|c|c|c|c|c|c|c|}\hline\hline
                       &$A_1(0)$            &$A_{+}(0)$            &$A_{-}(0)$            &$V(0)$            &$F_{+}(0)$            &$F_{-}(0)$   \\
                       &[$m_{fit}$]         &[$m_{fit}$]           &[$m_{fit}$]           &[$m_{fit}$]       &[$m_{fit}$]           &[$m_{fit}$] \\ \hline
 QCDSR \cite{ff-QCDSR} &0.63                &0.69                  &$-1.12$               &1.03              &0.66                  &$-0.36$   \\
                       &[4.5]               &[4.5]                 &[4.5]                 &[4.5]             &[4.5]                 &[4.5] \\ \hline
 RQM1 \cite{ff-RQM1}   &0.56                &0.54                  &$-0.95$               &0.83              &0.61                  &$-0.32$  \\
                       &[5.45]              &[4.76]                &[4.68]                &[4.72]            &[4.84]                &[4.80]\\ \hline\hline

                       &$A_1(0)$            &$A_{2}(0)$            &$A_{0}(0)$            &$V(0)$            &$F_{1}(0)$            &$F_{0}(0)$   \\
                       &[$m_{fit}$]         &[$m_{fit}$]           &[$m_{fit}$]           &[$m_{fit}$]       &[$m_{fit}$]           &[$m_{fit}$] \\ \hline
 RQM2 \cite{ff-RQM2}   &0.50                &0.73                  &0.40                  &0.49              &0.47                  &0.47      \\
                       &[4.84]              &[4.72]                &[4.04]                &[3.99]            &[4.41]                &[4.72]    \\ \hline\hline

                       &$A_1(0)$            &$A_{2}(0)$            &$A_{0}(0)$            &$V(0)$            &$F_{1}(0)$            &$F_{0}(0)$   \\
                       &[$c_1/c_2$]         &[$c_1/c_2$]           &[$c_1/c_2$]           &[$c_1/c_2$]       &[$c_1/c_2$]           &[$c_1/c_2$] \\ \hline
 LFQM \cite{ff-LFQM}   &0.50                &0.44                  &0.53                  &0.74              &0.61                  &0.61  \\
                       &[1.73/0.33]         &[2.22/0.45]           &[2.39/0.50]           &[2.46/0.56]       &[1.99/0.44]           &[1.18/0.17]\\ \hline
 \hline
\end{tabular}
\end{center}
\caption{ The parameters for the weak form-factors, where the unit of the $m_{fit}$ is GeV. }
\end{table}

   The numerical values of the branching fractions are shown in Table 2, from the table, we can see that the branching fractions vary in a large range according  to the values of the weak form-factors from different theoretical approaches, it is difficult to   determine which one is superior to others.  In Table 2, we also present the predictions from the Berezhnoy-Likhoded-Luchinsky (BLL) model  for comparison, where the spectral function
    \begin{eqnarray}
\rho_T(q^2)&=&5.86 \times10^{-5}\left(1-\frac{9m^2_{\pi}}{q^2}\right)^4\frac{1+190q^2}{\left[ (q^2-1.06)^2+0.48 \right]^2}
\end{eqnarray}
     determined from the $\tau\to\nu_{\tau} \pi^+\pi^-\pi^+$ decays is used \cite{Likhoded-3pi,Berezhnoy-3pi,Berezhnoy-3pi-2}. The present values are slightly
     different from that of Ref.\cite{Likhoded-3pi}, as we have taken slightly different  input parameters.

       The ratios among the branching fractions are shown explicitly in Table 3, from the table, we can see that
\begin{eqnarray}
\frac{Br(B_c^+ \to J/\psi \pi^+\pi^-\pi^+)}{Br(B_c^+ \to J/\psi \pi^+)} &=&  2.17\, , \, 1.87\, , \, 2.28\, , \, 1.88\, ,
\end{eqnarray}
are all compatible with the experimental data $2.41\pm0.30\pm0.33$ within uncertainties \cite{LHCb-Bc}, while the ratios based on the weak form-factors from the QCD sum rules in Ref.\cite{ff-QCDSR} and the relativistic quark models in Ref.\cite{ff-RQM2} are better. All those predictions can be confronted with the experimental data in the future at the LHCb collaboration.

In Fig.1, we plot the differential  decay widths of the $B_c$ mesons  $d\Gamma(B_c^+ \to J/\psi\, \pi^+\pi^-\pi^+)/dq^2$,
        $d\Gamma(B_c^+ \to \eta_c \,\pi^+\pi^-\pi^+)/dq^2$, $d\Gamma(B_c^+ \to J/\psi \, \pi^+\pi^-\pi^+)/dl^2$ and $d\Gamma(B_c^+ \to \eta_c \, \pi^+\pi^-\pi^+)/dl^2$ with variations of the squared  momenta $q^2$ ($M^2(\pi^+\pi^-\pi^+)$) and $l^2$ ($M^2(\pi^+\pi^-)$). By measuring the momenta dependence of the differential  decay widths, we can test dominance  of the intermediate axial-vector meson $a_1(1260)$ and   vector meson $\rho(770)$ in the invariant $\pi^+\pi^-\pi^+$ and $\pi^+\pi^-$  mass distributions, respectively. It is difficult to distinguish the two $\pi^+$ mesons in the final states, so we should not be serious for taking the  squared virtual momentum $l^2$ as the invariant mass distribution $M^2(\pi^+\pi^-)$. However, from Fig.2 we can see that the approximation $l^2=M^2(\pi^+\pi^-)$ works rather well.

In Fig.2, we plot the $\pi^+\pi^-\pi^+$ and $\pi^+\pi^-$  invariant mass distributions  in the decays $B_c \to J/\psi\pi^+\pi^-\pi^+$ compared with
the experimental data and the predictions of the BLL model. With suitable normalization,  the weak form-factors from  the QCD sum rules  \cite{ff-QCDSR}, the relativistic quark models \cite{ff-RQM1,ff-RQM2},   and the light-front quark models  \cite{ff-LFQM} lead to almost the same line-shapes for the invariant mass distributions, although they correspond to quite different decay widths.  From the figure, we can see that  the present work and the BLL model both  describe the experimental data on the invariant mass distributions $M(\pi^+\pi^-\pi^+)$ well,  while the present prediction is better.

\begin{table}
\begin{center}
\begin{tabular}{|c|c|c|c|c|c|c|}\hline\hline
                                             & QCDSR \cite{ff-QCDSR}  & RQM1 \cite{ff-RQM1} & RQM2 \cite{ff-RQM2}  & LFQM \cite{ff-LFQM}  \\ \hline
           $B^+_{c}\to J/\psi\, \pi^+$       & 1.222                  & 1.106               & 0.544                & 0.956         \\ \hline
           $B^+_{c}\to \eta_c\, \pi^+$       & 1.592                  & 1.360               & 0.879                & 1.360         \\ \hline
 $B^+_{c}\to J/\psi\, \pi^+\pi^-\pi^+$       & 2.655 [3.475]          & 2.072 [2.758]       & 1.242 [1.607]        & 1.797 [2.390]  \\ \hline
 $B^+_{c}\to \eta_c\, \pi^+\pi^-\pi^+$       & 1.854                  & 1.544               & 0.947                & 1.578        \\ \hline
 \hline
\end{tabular}
\end{center}
\caption{ The branching fractions of the $B_c$ decays, where the unit is $10^{-3}$, and   the references denote the hadronic form-factors from that
   articles  are used.  The values in the bracket denote the predictions  from the BLL model, where the spectral function
   $\rho_T(q^2)$ determined from the $\tau\to\nu_{\tau} \pi^+\pi^-\pi^+$ decays is used.}
\end{table}

\begin{table}
\begin{center}
\begin{tabular}{|c|c|c|c|c|c|c|}\hline\hline
                                                                      & QCDSR \cite{ff-QCDSR} & RQM1 \cite{ff-RQM1} & RQM2 \cite{ff-RQM2} & LFQM \cite{ff-LFQM} \\ \hline
$\frac{Br(B_c^+\to J/\psi\pi^+\pi^-\pi^+)}{Br(B_c^+\to J/\psi\pi^+)}$ & 2.17                  & 1.87                & 2.28                & 1.88  \\ \hline
$\frac{Br(B_c^+\to\eta_c\pi^+\pi^-\pi^+)}{Br(B_c^+\to J/\psi\pi^+)}$  & 1.52                  & 1.40                & 1.74                & 1.65  \\ \hline
$\frac{Br(B_c^+\to\eta_c \pi^+)}{Br(B_c^+\to J/\psi \pi^+)}$          & 1.30                  & 1.23                & 1.62                & 1.42   \\ \hline
$\frac{Br(B_c^+\to\eta_c\pi^+\pi^-\pi^+)}{Br(B_c^+\to \eta_c \pi^+)}$ & 1.16                  & 1.14                & 1.08                & 1.16   \\ \hline
 \hline
\end{tabular}
\end{center}
\caption{ The ratios among the branching fractions of the $B_c$ decays, where   the references denote the hadronic form-factors from that
   articles  are used. }
\end{table}

\begin{figure}
 \centering
 \includegraphics[totalheight=5cm,width=6cm]{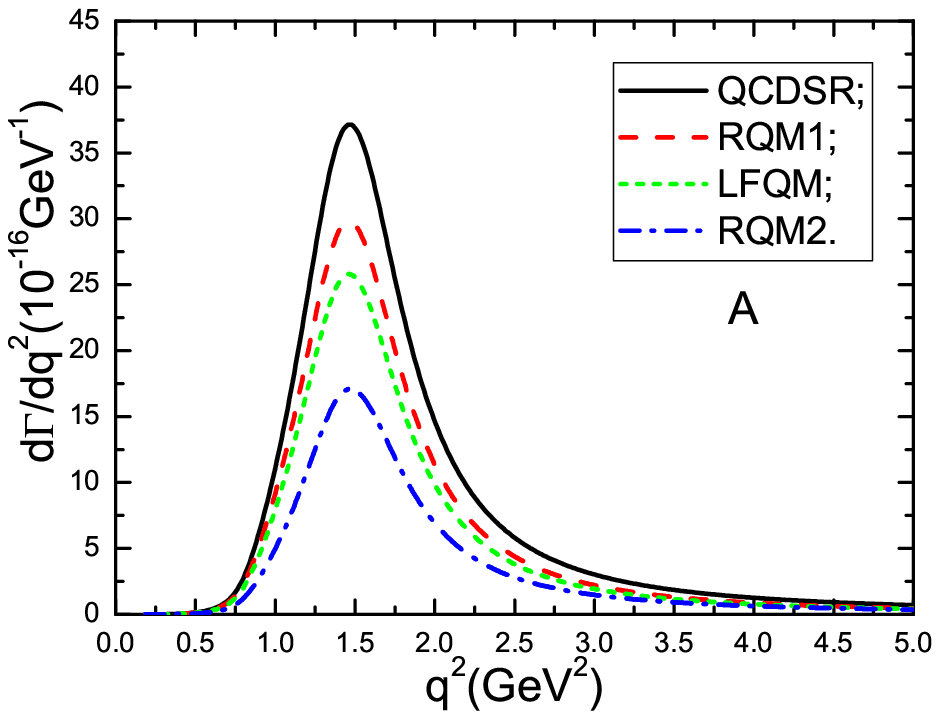}
\includegraphics[totalheight=5cm,width=6cm]{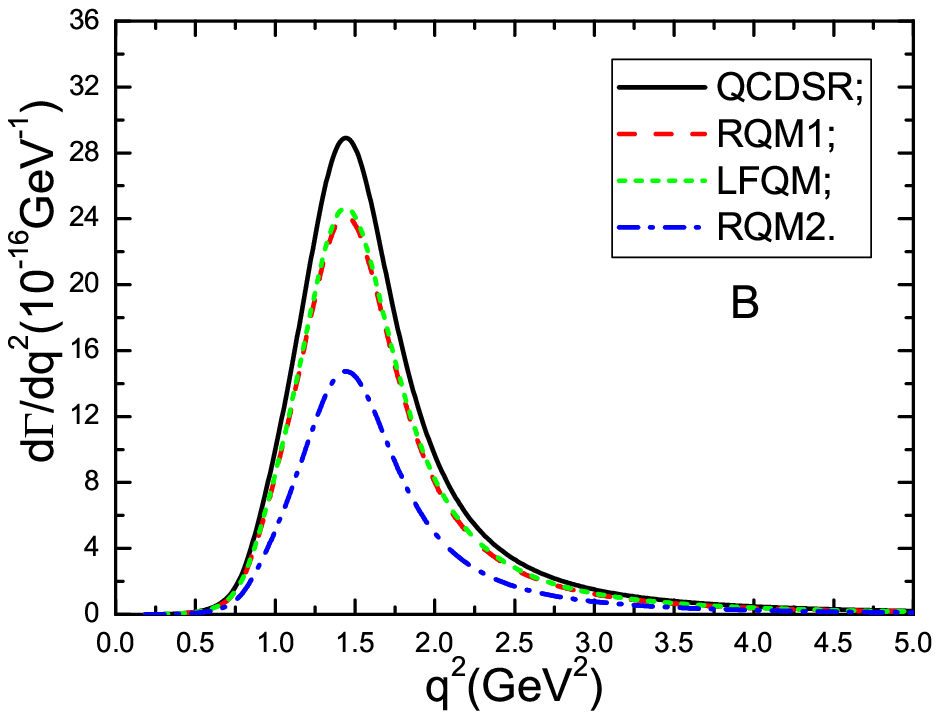}
\includegraphics[totalheight=5cm,width=6cm]{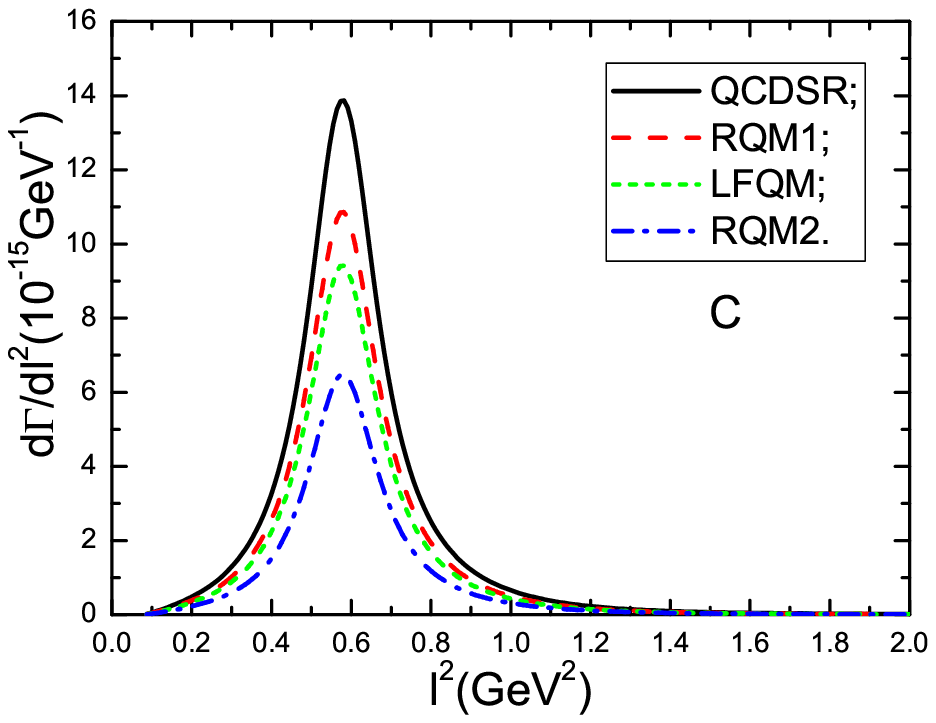}
\includegraphics[totalheight=5cm,width=6cm]{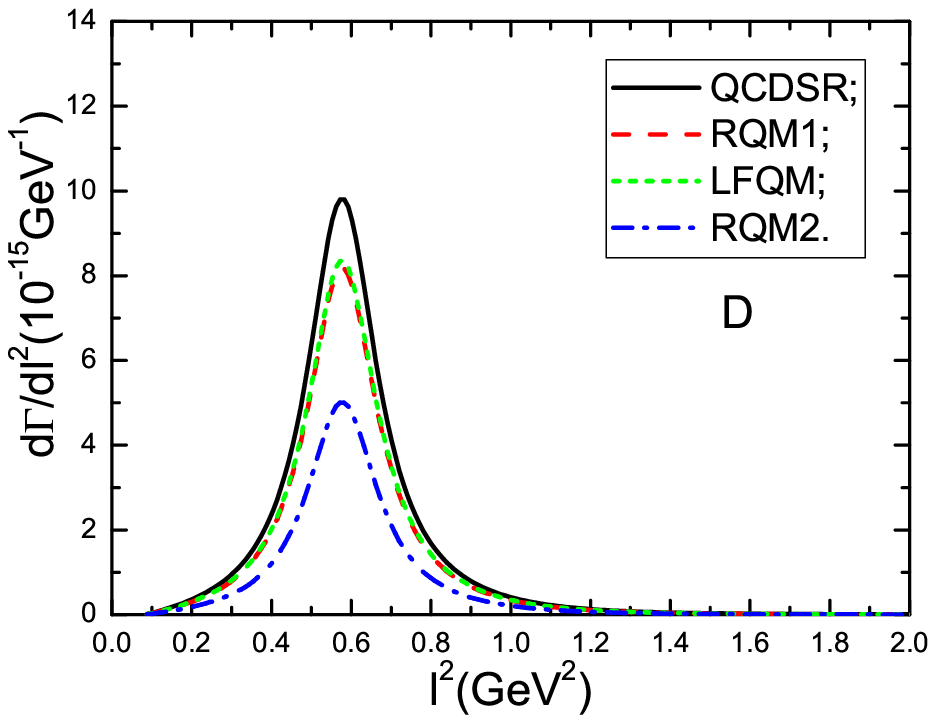}
        \caption{ The differential  decay widths of the $B_c$ meson, where the $A$, $B$, $C$ and $D$ denote the differential  decay widths $d\Gamma(B_c^+ \to J/\psi \pi^+\pi^-\pi^+)/dq^2$,   $d\Gamma(B_c^+ \to \eta_c \pi^+\pi^-\pi^+)/dq^2$, $d\Gamma(B_c^+ \to J/\psi \pi^+\pi^-\pi^+)/dl^2$ and $d\Gamma(B_c^+ \to \eta_c \pi^+\pi^-\pi^+)/dl^2$, respectively.}
\end{figure}

\begin{figure}
 \centering
 \includegraphics[totalheight=5cm,width=6cm]{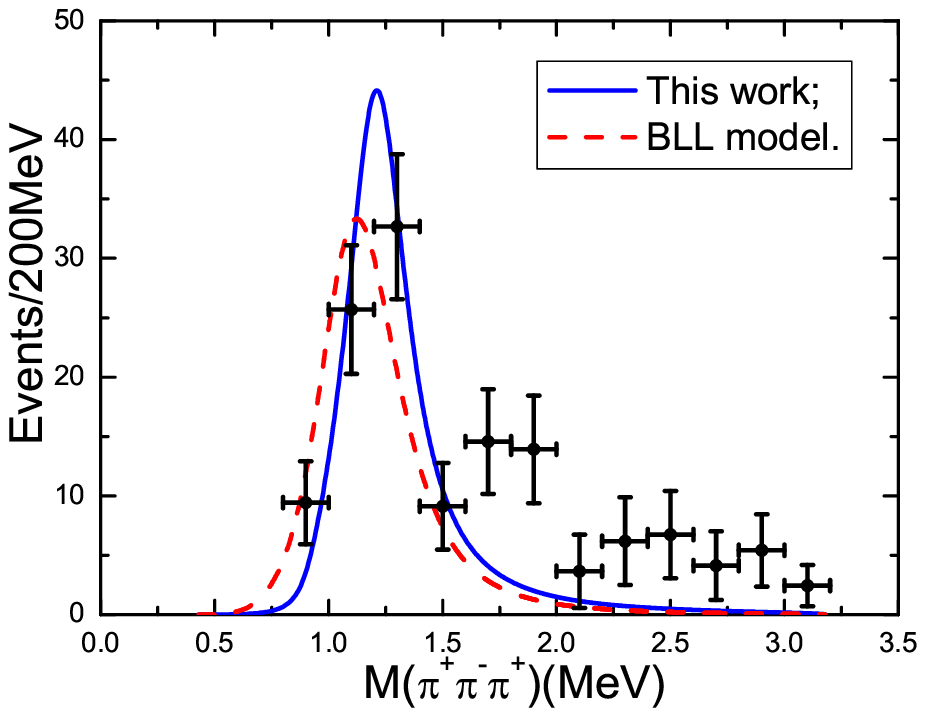}
\includegraphics[totalheight=5cm,width=6cm]{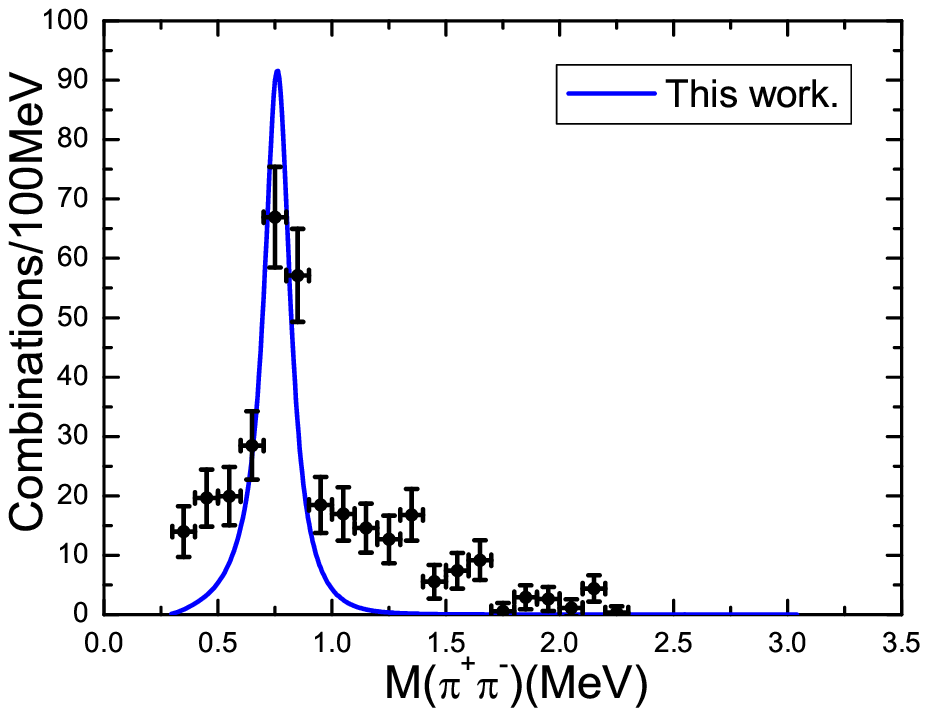}
        \caption{ The invariant mass distributions of the $\pi^+\pi^-\pi^+$ and $\pi^+\pi^-$ compared with the experimental data, where we have taken the approximation $l^2=M^2(\pi^+\pi^-)$.}
\end{figure}

\section{Conclusion}
In this article, we study the three-pion $B_c$-decays $B_c^+ \to J/\psi\, \pi^+\pi^-\pi^+$, $\eta_c\, \pi^+\pi^-\pi^+ $ assuming   dominance of the intermediate  axial-vector   meson $a_1(1260)$ and vector meson $\rho(770)$   in the invariant $\pi^+\pi^-\pi^+$ and $\pi^+\pi^-$ mass distributions  respectively, and make predictions for the branching fractions and differential decay widths. The  ratios between the decays $B_c^+ \to J/\psi\, \pi^+\pi^-\pi^+$ and $B_c^+ \to J/\psi\, \pi^+ $  based on the form-factors from different theoretical approaches are compatible with the  experimental data within uncertainties. The decays $B_c^+ \to \eta_c\, \pi^+\pi^-\pi^+$, $\eta_c\, \pi^+ $ have not been observed yet, the predictions can be confronted with the experimental data in the future at the LHCb.

\section*{Acknowledgements}
This  work is supported by National Natural Science Foundation,
Grant Number 11075053,  and the Fundamental Research Funds for the
Central Universities.


\begin{thebibliography}{99}

\bibitem{CDF1998}   F. Abe  et al,  Phys. Rev. Lett. {\bf 81} (1998) 2432;
F. Abe et al, Phys. Rev. {\bf D58} (1998) 112004.

\bibitem{GI} S. Godfrey and N. Isgur, Phys. Rev. {\bf D32} (1985)  189; S. Godfrey, Phys. Rev. {\bf D70} (2004) 054017.

\bibitem{CDF-life} A. Abulencia et al, Phys. Rev. Lett. {\bf 97} (2006) 012002.

\bibitem{D0-life}  V. Abazov et al, Phys. Rev. Lett. {\bf 102} (2009) 092001.

\bibitem{CDF2008} T. Aaltonen et al, Phys. Rev. Lett. {\bf 100} (2008) 182002.

\bibitem{D02008}  V. M. Abazov et al, Phys. Rev. Lett. {\bf 101} (2008) 012001.

\bibitem{PDG} K. Nakamura et al, J. Phys. {\bf G37} (2010) 075021.

\bibitem{LHCb-Bc}  R. Aaij et al,  Phys. Rev. Lett. {\bf 108} (2012) 251802.

\bibitem{Likhoded-3pi} A. K. Likhoded and A. V. Luchinsky, Phys. Rev. {\bf D81} (2010) 014015.

\bibitem{Berezhnoy-3pi} A. V. Berezhnoy, A. K. Likhoded and A. V. Luchinsky, arXiv:1104.0808.

\bibitem{Berezhnoy-3pi-2} A. V. Berezhnoy, A. K. Likhoded and A. V. Luchinsky, PoS QFTHEP2011 (2011) 076.


\bibitem{Rakitin-3pi} A. Rakitin and S. Koshkarev, Phys. Rev. {\bf D81} (2010) 014005.

\bibitem{Kuhn-3pi-ZPC}  J. H. Kuhn and A. Santamaria, Z. Phys. {\bf C48} (1990) 445.

\bibitem{Li-3pi-PRD}  B. A. Li, Phys. Rev. {\bf D55} (1997) 1436.

\bibitem{Ivanov-3pi-ZPC} Y. P. Ivanov, A. A. Osipov and M. K. Volkov, Z. Phys. {\bf C49} (1991) 563.

\bibitem{Kuhn-3pi-PLB} J. H. Kuhn and E. Mirkes, Phys. Lett. {\bf B286} (1992) 381.

 \bibitem{a1-rho-pi}  P. Lichard and J. Juran, Phys. Rev. {\bf D76} (2007) 094030.

 \bibitem{rho-pi-pi} H. Y. Cheng, C. K. Chua and A. Soni,  Phys. Rev. {\bf D71} (2005) 014030.


 \bibitem{ff-QCDSR} V. V. Kiselev, hep-ph/0211021.

 \bibitem{WSB} M. Wirbel, B. Stech and M. Bauer,  Z. Phys. {\bf C29} (1985) 637.

 \bibitem{KLO-QCDSR} V. V. Kiselev, A. K. Likhoded and A. I. Onishchenko,  Nucl. Phys. {\bf B569} (2000) 473.

\bibitem{Colangelo-QCDSR}  P. Colangelo, G. Nardulli and N.  Paver,   Z. Phys. {\bf C57} (1993) 43.

\bibitem{HuangZuo} T. Huang and F. Zuo, Eur. Phys. J. {\bf C51} (2007) 833.

\bibitem{ff-RQM1} M. A. Ivanov, J. G. Korner and P. Santorelli, Phys. Rev. {\bf D71} (2005) 094006.

\bibitem{ff-RQM2} D. Ebert, R. N. Faustov and V. O. Galkin, Phys. Rev. {\bf D68} (2003) 094020.

\bibitem{IKS-RQM} M. A. Ivanov, J. G. Korner and P. Santorelli, Phys. Rev. {\bf D63} (2001) 074010.

\bibitem{Nobes-RQM} M. A. Nobes and R. M. Woloshyn, J. Phys. {\bf G26} (2000) 1079.

\bibitem{ff-LFQM} W. Wang, Y. L. Shen and C. D. Lu, Phys. Rev. {\bf D79} (2009) 054012.

\bibitem{Choi-LFQM} H. M. Choi and C. R. Ji, Phys. Rev. {\bf D80} (2009) 054016.

\bibitem{Hernandez-NRQM}  E. Hernandez, J. Nieves and J.M. Verde-Velasco, Phys. Rev. {\bf D74} (2006) 074008.

\bibitem{Dhir-NRQM} R. Dhir and R. C. Verma, Phys. Rev. {\bf D79} (2009) 034004.

\bibitem{Sun-pQCD} J. F. Sun, D. S.  Du and Y. L. Yang, Eur. Phys. J. {\bf C60} (2009) 107.

\bibitem{Compass}   M. G. Alekseev et al, Phys. Rev. Lett. {\bf 104} (2010) 241803.


\bibitem{Yang-fa} K. C. Yang, Nucl. Phys. {\bf B776} (2007) 187.


\end{thebibliography}
\end{document}